\documentclass[a4paper, 11pt]{article}

\usepackage{amssymb}
\usepackage{latexsym}
\usepackage{amsmath}

\newcommand{\sgn}{\,{\rm sgn}\,}
\def\be{\begin{equation}}
\def\ee{\end{equation}}
\def\Res{{\rm Res}}
\def\half{ \frac{1}{2}}

\begin{document}

  \begin{flushright}
    hep-th/0407035\\
    DAMTP-2004-71    
  \end{flushright}
  \vskip 1cm
  \begin{center}
    \LARGE
   {\bf  A note on the uniqueness of the Neumann matrices} \\ {\bf in the
    plane-wave background }   
  \end{center}
  \vskip 0.5cm
  \begin{center}
        \begin{center}
        {\Large James Lucietti }\\
        \bigskip\medskip
        {\it  DAMTP, Centre for Mathematical Sciences,\\
	  University of Cambridge,
	  Wilberforce Rd.,\\
	  Cambridge CB3 0WA, UK\\}
        {\small E-mail: {\tt J.Lucietti@damtp.cam.ac.uk} }
        \end{center}
  \end{center}
  \vskip 0.5cm
\begin{center}
{\bf Abstract}
\end{center}
\centerline{
\parbox[t]{15cm}{\small
\noindent
In this note, we prove the uniqueness of the Neumann matrices of the open-closed
 vertex in plane-wave light-cone string-field theory, first derived
  for all values of the mass parameter $\mu$ in~\cite{lss}. We also prove the existence and uniqueness of the
 inverse of an infinite dimensional matrix  necessary for the cubic vertex Neumann matrices, and give an explicit
 expression for it in terms of $\mu$-deformed Gamma functions. Methods of complex
 analysis are used together with the analytic properties of the
 $\mu$-deformed Gamma functions. One of the implications of these
 results is that the geometrical continuity
conditions suffice
to determine the bosonic part of the vertices as in flat space.}}

\vskip1.2cm
The plane-wave limit of the AdS/CFT correspondence has provided a
concrete arena for testing the validity of the duality. In this limit,
we have a relation between string theory in the plane-wave limit of
$AdS_5 \times S^5$~\cite{metsaev, metsaevtseytlin} and a certain sector (BMN) of ${\cal N}=4$,
$d=4$ super Yang-Mills (SYM) theory~\cite{bmn}. The beauty of this limit is that
both the string theory and gauge theory side are perturbative. 

Interestingly, this area has also motivated the development of certain deformations of classical special functions such as the
Theta functions and the Gamma function. The deformed Theta functions appeared in expressions for the cylinder diagrams that determine the
static interactions between pairs of $Dp$-branes in the type IIB
plane-wave background~\cite{Bergman, gg, GGSS}. The so-called
``$\mu$-deformed Gamma functions'' first appeared,
in several flavours, in the expressions for the Neumann matrices of
the open-closed vertex~\cite{lss} and the cubic vertex of plane-wave light-cone
string field theory~\cite{lsscubic}. These Neumann matrices beautifully generalise the
ones in flat-space obtained long ago~\cite{GSstringfield, GSstringint} in terms of the ordinary Gamma
function. Other unexpected results in classical complex analysis have
occurred from research in this area, such as a generalisation of an
integral transform called the Stieltjes transform~\cite{schwarz}, which came about
from the original derivation of the cubic vertex~\cite{hssv, schwarzreview}.
\\ 

In this letter, we provide a proof of the uniqueness of the Neumann
matrices $A_{mn}$, $B_{mn}$ and $C_{mn}$, which were determined in the
solution to the bosonic part of the open-closed vertex~\cite{lss}. We
also provide a proof of existence and uniqueness of the inverse of a certain infinite
dimensional matrix called $\Gamma_+$, from
which we deduce uniqueness of the cubic vertex Neumann matrices
$\bar{N}^{rs}_{mn}$, which were determined in~\cite{hssv, lsscubic}.
\\

As a reminder, in~\cite{lss} we solved\footnote{We should note that
  the large $\mu$ asymptotics were first derived in~\cite{Gomis}
  without the knowledge of the exact expression valid for all $\mu$
  first obtained in~\cite{lss}.} for the bosonic part of the
vertex $|V \rangle_B$ using the geometrical continuity conditions
\be
X^i(\sigma)_{open}|V \rangle_B = X^i(\sigma)_{closed} |V \rangle_B\,, \qquad 
[P^i(\sigma)_{open} + P^i(\sigma)_{closed}]|V \rangle_B=0 \,, 
\ee
which are understood to hold at $\tau=0$, the interaction time
($i=1,...,8$ are the transverse coordinates). Working
in terms of modes allows one to deduce an expression for the vertex
entirely analogous to the one in flat space\footnote{We will only concern
ourselves with the case of Neumann boundary conditions for the open
string. In~\cite{lss} Dirichlet boundary conditions were also addressed.}
\be
|V \rangle_B = \exp( \Delta) |0 \rangle
\ee
where
\begin{eqnarray}
\Delta = -\sum_{m=1}^\infty {\sqrt{2}\over \omega_{2m}} \beta_{-2m}
\alpha^I_{-m} \nonumber  +\sum_{m,n=0}^\infty A_{mn} \beta_{-2m-1} \alpha^{II}_{-n} + \half B_{mn} \beta_{-2m-1}\beta_{-2n-1} + \half C_{mn} \alpha_{-m}^{II} \alpha_{-n}^{II} \,
\end{eqnarray}
and $\alpha^{I/II}_n  = \sqrt{2}(\alpha_n \pm \tilde{\alpha}_n)$,
$\omega_n = \sgn(n) \sqrt{n^2+\mu^2}$ and here $|0 \rangle$ is the
vacuum of the two
string Fock space. The modes
$\alpha_n$ and $\tilde{\alpha}_n$ are of the closed string, and
$\beta_m$ of the open string, see~\cite{lss} for the explicit
modes expansions and commutators.
The Neumann matrices $A_{mn}$, $B_{mn}$ and $C_{mn}$ satisfy the following
complicated set of coupled equations:
\begin{eqnarray}
&&-{2\sqrt{2} i n\over \pi} \sum_{m=0}^{\infty}
{1\over 
\omega_{2m+1}- \omega_{2n}}
A_{mk} = \delta_{n,k} \\
&&-\sum_{m=0}^\infty {B_{mk}\over \omega_{2m+1} - \omega_{2n}}
= {1\over (\omega_{2n} +\omega_{2k+1})\omega_{2k+1}} \\
&&{4\sqrt{2} i n\over \pi \omega_{2n}} \sum_{m=0}^{\infty} { 1\over
  \omega_{2m+1}+ \omega_{2n}} A_{mp}= C_{np} \label{C} \\
&&{4\sqrt{2} i n\over \pi\omega_{2n}} \left(\sum_{m=0}^\infty {1\over \omega_{2m+1}+ \omega_{2n}}
B_{mp} + {1\over (\omega_{2p+1}-\omega_{2n}) \omega_{2p+1}}\right)= \,
A_{pn}\,.
\end{eqnarray}
We found that a solution to this system of equations is given by
\begin{eqnarray}
A_{mk} &=& i\sqrt{2}  {v^I_m v^{II}_k \over (\omega_{2m+1} -
\omega_{2k})} \\
B_{mk} &=& {v^I_m v^I_k \over (\omega_{2m+1} +
\omega_{2k+1})} \\
C_{mk} &=& 2 { v^{II}_m v^{II}_k \over (\omega_{2m} + \omega_{2k}) }
\,.
\end{eqnarray}
These are written in terms of deformations of the binomial
coefficients $u_n={\Gamma (n+1/2) \over \sqrt{\pi}\, \Gamma (n+1)} $
of $(1-x)^{-1/2} = \sum_{n=0}^{\infty} { u_n x^n \over n!}$, which take
the form
\begin{eqnarray}
v_m^I &=& {(2m+1) \over \omega_{2m+1}} \  
{\Gamma^I_{\mu} (m+1/2) \over \sqrt{\pi}\; \Gamma^{II}_{\mu}(m+1)} \\
v_m^{II} &=& {2\over \omega_{2m}}\  { \Gamma^{II}_{\mu}(m+1/2) \over \sqrt{\pi}\;
\Gamma^I_{\mu} (m)} \,,
\end{eqnarray}
where $\Gamma^I_{\mu}(z)$ and $\Gamma^{II}_{\mu}(z)$ are the
$\mu$-deformed Gamma functions of the first and second kind introduced
in~\cite{lss}, which
both reduce to $\Gamma(z)$ in the flat space limit $\mu \to 0$. One
can find the definitions of these functions, together with some of
their key properties in the Appendix.

In that paper
a method was suggested for proving uniqueness of $A_{mn}$, $B_{mn}$ and
$C_{mn}$, at least in flat space,
however this relied on some fairly heavy-handed complex analysis. It
first involved determining the singularity structure of the matrices.
This is the weak point of the technique, as this could only really
be motivated, albeit strongly, and not proved. The method used here however, in light of the
complicated nature of the set of equations which the Neumann matrices
satisfy, is remarkably simple. It relies on an elementary observation
of the property of inverses of infinite dimensional matrices. Namely,
if both a right and left inverse exist then they are equal and
{\it unique}. Note that for infinite dimensional matrices one can have the
situation where we have multiple left inverses but no right inverses,
for example (see Appendix D of~\cite{GSstringint}). Fortunately it so happens that one of the Neumann
matrices, $A_{mn}$, is actually the right inverse of a certain
infinite dimensional matrix. Thus a proof of uniqueness rests on
showing that it is also the left inverse, which we prove in
this letter.
\\

In this letter we also prove existence and uniqueness of the Neumann
matrices $\bar{N}^{rs}_{mn}$ of the cubic vertex. As a
reminder~\cite{Spradlin1, Spradlin2} the
bosonic part of the vertex is written as 
\be
| V \rangle_B = \exp \left( \frac{1}{2}
\sum_{r,s=1}^{3}\sum_{m,n=-\infty}^{\infty}\sum_{i=1}^{8}
a_{rm}^{i\dagger}\bar{N}_{mn}^{rs} a^{i\dagger}_{sm} \right)|0 \rangle \,
\ee
where $a^i_{rm}$ are the modes of the three closed strings, normalised
to satisfy the harmonic oscillator algebra $[ a^i_{rm},
  a_{sn}^{j\dagger} ] = \delta_{mn}\delta^{ij}$ and $m \in
\mathbb{Z}$ (note $a^{\dagger}_m \neq a_{-m}$ here). Here, the state $|0
\rangle$ denotes the three string Fock vacuum (this satisfies $a^i_{rm}| 0
\rangle=0$ for  $m \in \mathbb{Z}$). The three strings
have incoming momenta $\alpha_{r} \equiv 2 {p^{(r)}}^+$, so
$\sum_{r=1}^{3} \alpha_{r}=0$.
As in flat space we can ``factorise'' the Neumann matrices in
terms of Neumann vectors $\bar{N}^r_m$~\cite{Schwarzcomments, Panki}, such
that for $m,n \geq 1$ \footnote{One can relate the Neumann matrices for
  the negative mode numbers to the ones with positive mode numbers, so
  it is sufficient to restrict to positive ones. In particular, the
  only non-vanishing matrix elements for negative modes are given by
  $\bar{N}^{rs}_{-m-n}= -(U^{(r)}\bar{N}^{rs}U^{(s)})_{mn}$.}
\be
\bar{N}^{rs}_{mn}= -\frac{mn\alpha}{1+\mu\alpha k} \frac{\bar{N}^r_m
  \bar{N}^s_n}{\alpha_s\omega_{r,m}+\alpha_r \omega_{s,n}},
\ee
where $\omega_{r,m} = \sgn{(n)}\sqrt{n^2 +\alpha_r^2\mu^2}$.
The Neumann vectors can be written as
\be
\bar{N}^r_m = \sqrt{\frac{\omega_{r,m}}{m}}
\frac{(\omega_{r,m}+\alpha_r\mu)}{m} \frac{1}{\alpha_r} f^{(r)}_m
\ee
 and in~\cite{lsscubic} we determined the vectors $f^{(r)}_m$, and the
 scalar $k=B^Tf^{(3)}$, by
 finding a solution to the two sums
\begin{eqnarray}
\label{sum1}
\sum_{p=1}^\infty \, f^{(3)}_p  A^{(r)}_{pm}  &=& {\alpha_3\over \alpha_r}
\,  f^{(r)}_m  \\ \label{sum2}
\sum_{p=1}^\infty \sum_{r=1}^3 {1\over \alpha_r} \, \left(A^{(r)}
U^{(r)}\right)_{mp} f^{(r)}_p &=&- B_m\,
\end{eqnarray}
which when combined give $\Gamma_+ f^{(3)} = B$; these sums can be
derived from the appropriate geometrical continuity conditions. See
the Appendix for the definitions of the matrices $A^{(r)}$, $U^{(r)}$
and the vector $B$. The solutions we
found are expressed in terms of $\mu$-deformed Gamma functions as follows:
\be
f^{(r)}_m= { e^{\tau_0 (\mu+\omega_{{m\over \alpha_r}})}
\over \sqrt{m} (-\alpha_{r} -\alpha_{r+1}) \, 
\omega_{m\over \alpha_r}} 
{\Gamma_{\mu}^{(r+1)}\left(- {m\over \alpha_r}\right) \over 
\Gamma_{\mu}^{(r)}\left({m\over \alpha_r}\right)
\Gamma_{\mu}^{(r-1)} \left({m\over \alpha_r}\right)} 
\, M(0^+)\, ,
\ee
and 
\be 
k= \frac{1}{\alpha \mu}( M(0^+)^2e^{2\tau_0\mu}-1) \, .
\ee
See the Appendix for definitions of the Gamma functions and the
constant $M(0^+)$.
Note that an alternate, less direct method was first used
in~\cite{hssv} to derive these quantities. To relate the notations of
the two papers~\cite{hssv, lsscubic}, we have $Y=f^{(3)}$ and $k=-4K$, although for
convenience here we will use $k$.

One might ask whether a simple proof of uniqueness could be presented for the
Neumann matrices of the cubic vertex as well. As stressed, in~\cite{lsscubic} we found a
solution $f^{(3)}$ to $\Gamma_+f^{(3)} = B$. It is clear that if we could show
that $\Gamma_+$ possesses a left inverse then the solution $f^{(3)}$ is
indeed unique. At this point we note a welcome simplification: since
$\Gamma_+$ is actually a symmetric matrix, existence of a left inverse
implies existence of a right inverse, which in turn tells us that
these inverses are equal and unique. Thus the problem of uniqueness in
this case reduces to showing that $\Gamma_+^{-1}$ actually exists!
 It appears that no-one has actually checked existence of
this matrix. In~\cite{hssv} an expression for the inverse is derived\footnote{More precisely an expression for the vector $Y$ and scalar $k$
is derived which is enough information to deduce $\Gamma_+^{-1}$ using
an identity derived in~\cite{Schwarzcomments}.}
{\it assuming} it exists - of course to complete the proof one needs
to verify that the expression is the inverse directly. This is
possible using the methods of~\cite{lss, lsscubic}. We will present
this calculation too, however it is rather more involved than than the
calculation for the open-closed vertex. The great advantage of using the deformed Gamma
functions in the expressions of the Neumann matrices is that the
infinite dimensional matrix algebra required can be done explicitly
just using elementary complex analysis. In fact what we will do is
evaluate the infinite sum $\Gamma_+ \Gamma_+^{-1}$ and confirm is it
equal to the identity.
\\

A proof of uniqueness is important in the plane-wave theory. It means
that the geometrical continuity conditions imposed to derive the set of
equations satisfied by the Neumann matrices, using Fock space methods, actually suffice to
settle the bosonic part of the vertex. This is a nice result as it
means that this method can probably be used more generally. For example, one
might contemplate using geometrical continuity and the oscillator
methods to compute the 1-loop correction to the closed string
propagator, rather than directly ``gluing'' two cubic vertices together. Note that in flat-space one
could deduce uniqueness by appealing to the alternate Green's function
method of deriving the matrices. However, these methods, at the present
at least, are absent in the plane-wave case due to the lack of
explicit conformal invariance on the worldsheet. The proofs given
here actually work in flat-space too (as they are valid for all
$\mu$!) and it appears that they have not been noticed before.

Finally, we should emphasise the advantages of our method as compared to the
one used in~\cite{hssv}:
\begin{itemize}
\item It is not clear how one would
generalise the technique used in~\cite{hssv} to other interaction vertices, whereas the techniques
developed and used in~\cite{lss, lsscubic} are clearly more
generally applicable as they allow one to solve the geometrical
continuity conditions directly.

\item Our method is a direct generalisation
of the techniques which can be used in flat space.

\item Expressing the Neumann matrices in
terms of the deformed Gamma functions is not merely of aesthetic
value. Once armed with their analytic properties it provides a powerful
calculation tool.
\end{itemize}

Now we present our proofs.
As we have already hinted at, the proof is based on the following observation. Consider the equation
\be
\label{MA}
\sum_{m} M_{nm}A_{mk} = \delta_{nk} \; ,
\ee
which expresses the fact that $A$ is the right-inverse of $M$ when they
are interpreted as infinite dimensional matrices. If $A$ is also the
left-inverse to $M$ then the solution (i.e. solving for $A$ when $M$
is known) to (\ref{MA}) is unique. This follows since we have $\sum_k
A_{mk}M_{kn} = \delta_{mn}$, and thus if there exists another solution
to (\ref{MA}), say $A'$, then we have $A(MA')=A$ which implies $A'=A$ by
associativity~\footnote{Associativity of matrix multiplication in the
infinite dimensional case is equivalent to swapping the order of two
infinite sums and thus is related to the convergence properties of the
sums in question.}.

In solving for the open-closed vertex we came across the equation,
\be
\label{ocv}
\sum_{m=0}^{\infty} M_{nm} A_{mk} = \delta_{nk} \; ,
\ee
where 
\be
M_{nm} = { i2\sqrt{2} n \over \pi( \omega_{2n} - \omega_{2m+1}) } \; ,
\ee
and 
\be
A_{mk} = { i\sqrt{2} v^I_m v^{II}_k \over { \omega_{2m+1}
    -\omega_{2k}}} \; .
\ee
Therefore, based on the above observation if $A_{mn}$ is also the
left-inverse to $M_{mn}$ then it is the unique solution to (\ref{ocv}).
Thus we need to prove the following sum
\be
\label{AM}
\sum_{k=1}^{\infty} { A_{mk} 2\sqrt{2}i k \over
  \pi(\omega_{2k}-\omega_{2n+1})} = \delta_{mn} \; .
\ee
To this end consider the contour integral
\be
\label{int}
\oint {dk \over 2\pi i} \pi \cot(\pi k) A_{mk} { 2\sqrt{2}i k \over
  \pi (\omega_{2k} - \omega_{2n+1})} \; .
\ee
Recall $v^{II}_k = { 2\Gamma^{II}_{\mu}(k+1/2) \over \Gamma^I_{\mu}(k) \omega_{2k}
  \sqrt{\pi}}$, which tells us that $A_{mk}$ has zeroes for $k \in \mathbb{N}_0$,
simple poles at $k = -1/2, -3/2, -5/2,...$ and another simple pole at
$k=m+1/2$. $A_{mk}$ also has a branch cut on $[ i\mu/2 , -i\mu/2 ]$
and branch points at $\pm i\mu /2$. Therefore if $n \neq m$ the
integrand in (\ref{int}) only has poles at $k=1,2...$ (the rest are
cancelled), in which case the remaining contributions come from the
branch cuts, branch points and the integral at infinity. Using the
asymptotics of $v^{II}_k$ we see the integrand goes as $k^{-3/2}$ and
thus the integral at infinity does not contribute. Since $v^{II}_{0^{\pm}+iy}$ is
odd for $|y| < \mu /2$, the integrand on either side of the branch
cut is odd and thus these contributions vanish too. Finally the
integrals around the branch points vanish as $O(\epsilon^{1/2})$,
  where $\epsilon$ is the radius from the branch point. Thus
we have verified (\ref{AM}) for $n \neq m$. So now we consider $n=m$; the
argument proceeds in the same way except now the integrand also has a
simple pole at $k=n+1/2$. Thus we also need to include the following
residue
\begin{eqnarray}
\nonumber &&\Res_{k=n+1/2} \;  \pi \cot(\pi k) A_{nk} { 2\sqrt{2}i k \over
  \pi (\omega_{2k} - \omega_{2n+1})} \\ &&= 2(2n+1)v^I_{n+1/2} v^{II}_n
\lim_{k \to n+1/2} { \cot(\pi k) \over \omega_{2k} - \omega_{2n+1} }
\Res_{k=n+1/2} {1 \over \omega_{2k} - \omega_{2n+1}}  = -1 \; ,
\end{eqnarray}
which now proves (\ref{AM}). We deduce that $A_{mn}$ is unique and we are
entitled to call $M_{mn}=A^{-1}_{mn}$. 
Note it immediately follows that $C_{mn}$ is unique too, as it is
expressed entirely in terms of $A_{mn}$, see (\ref{C}). To prove that $B_{mn}$ is also unique,
we note that one of the equations satisfied by $B_{mn}$ is
\be
\label{sumB}
\sum_{m=0}^{\infty} M_{nm} B_{mk} = { 2\sqrt{2}in \over \pi
  \omega_{2k+1}(\omega_{2k+1}+\omega_{2n})} \;.
\ee
Therefore multiplying (\ref{sumB}) by $A_{pn}$ and summing from
$n=1,..., \infty$, and using (\ref{AM}) (i.e. using the fact that $A$ is left-inverse to $M$)  implies
\be
\label{B}
B_{mk} = \sum_{n=1}^{\infty} { 2\sqrt{2}i n A_{mn} \over \pi
  \omega_{2k+1}(\omega_{2n}+ \omega_{2k+1})} \; .
\ee
It follows that $B_{mn}$ is also unique, thus proving the
theorem. Note, one could of course check that the RHS of (\ref{B})
reproduces the correct $B_{mn}$ by doing the sum in the usual manner. Also
note the uniqueness theorem follows for flat space too by setting $\mu=0$.
\\

A crucial ingredient of the Neumann matrices of the cubic vertex, is the inverse of a
certain infinite dimensional matrix $\Gamma_+$. It is defined as 
\be
\Gamma_+ = \sum_{r=1}^{3} A^{(r)}U^{(r)}{A^{(r)}}^t
\ee
and one can find the definitions of the matrices $A^{(r)}$ and
$U^{(r)}$ in the Appendix. As already explained, there is a lose end
to tie up here. In~\cite{hssv}, the existence of the inverse was assumed
and using this an expression for it was derived. However no-one has checked
that the final explicit expression for $\Gamma_+^{-1}$ actually is the
inverse! In~\cite{lsscubic} we found a vector $f^{(3)}$ which
satisfied $\Gamma_+ f^{(3)} = B$. However we did not manage to prove it
was unique. Here we will settle both of these open ends by taking
the explicit expression for $\Gamma_+^{-1}$ in~\cite{hssv} and
multiplying it into $\Gamma_+$.  The existence of an inverse ensures
it is the unique left and right inverse (due to the matrix being
symmetric), and also establishes that the solution in~\cite{lsscubic} is unique.

Thus, the candidate expression for $\Gamma_+^{-1}$ is~\cite{Schwarzcomments}
\be
(\Gamma_+^{-1})_{mn} = \frac{m}{2\omega_{3,m}}\delta_{mn} +
\frac{\alpha_1\alpha_2(\omega_{3,m}+\mu\alpha_3)(\omega_{3,n}+\mu\alpha_3)
  f^{(3)}_mf^{(3)}_n}{2(\omega_{3,m}+\omega_{3,n})(1+\mu\alpha k)} \, .
\ee
 The sum we are interested in is
\be
\label{check}
\sum_{p=1}^{\infty}(\Gamma_+)_{mp}(\Gamma_+^{-1})_{pn}=
\frac{m}{2\omega_{3,m}}(\Gamma_+)_{mn} +
\frac{\alpha_1\alpha_2(\omega_{3,n}+\mu\alpha_3)f^{(3)}_n}{2(1+\alpha\mu
k)} \sum_{p=1}^{\infty} (\Gamma_+)_{mp} f^{(3)}_p
\frac{\omega_{3,p}+\mu\alpha_3}{\omega_{3,p}+\omega_{3,n}}.
\ee
Evaluation of the RHS of (\ref{check}) can be performed using two
identities which we will prove. We will work in the gauge
$\alpha_1=y$, $\alpha_2 =1-y$ and $\alpha_3=-1$. The first identity is:
\begin{eqnarray}
\label{lemma1}
\sum_{p=1}^{\infty}A^{(r)}_{pq} f^{(3)}_p \frac{\omega_p
  -\mu}{\omega_p +\omega_n} = \frac{\alpha_3}{\alpha_r} f^{(r)}_q
\frac{ \omega_{r,q}+\alpha_r\mu}{\omega_{r,q} - \alpha_r\omega_n} -
\frac{A^{(r)}_{nq}(\omega_n+\mu)(1+\mu\alpha k)}{\alpha_1\alpha_2 n
  \omega_n f^{(3)}_n}
\end{eqnarray}
which is only valid for $r=1,2$. For $r=3$ it is trivial to see we
only get the first term. Note that this sum is very similar to one
computed in~\cite{lsscubic}, namely (\ref{sum1}). Notice that the extra factor does not affect the
asymptotics of the integrand of the corresponding contour integral, or
the parity of the integrand along the branch cut. Thus we need only
worry about the extra residue which occurs at $p=-n$. This is what
gives the second term on the RHS of (\ref{lemma1}). The first term is the analogue of the RHS of (\ref{sum1}) which comes from the
residue at $p=-q/\alpha_r$. More explicitly, for the $r=1$ case, one needs to consider the
following contour integral
\begin{eqnarray}
\oint \frac{dp}{2\pi i}\frac{2}{\sin(\pi p)} \frac{(-1)^q \sqrt{q}
  M(0^+) y}{(1-y)} \frac{\sin(\pi
  yp)}{q^2-y^2p^2}\frac{(\omega_p-\mu)e^{\tau_0(\mu
  -\omega_p)}}{\omega_p(\omega_p+\omega_n)} \frac{ \Gamma_{\mu
  y}(yp)}{\Gamma_{\mu(1-y)}(-(1-y)p) \Gamma_{\mu}(p)}
\end{eqnarray}
where the contour is a large circle centred at the origin. The
integrand has simple poles for $p \in \mathbb{N}$, whose residues lead
to the sum we want, i.e. the LHS of (\ref{lemma1}); this is of course
by construction. Another simple pole occurs at $p=-q/y$ and leads to
the contribution $\frac{f^{(1)}_q}{y} \left(
\frac{\omega_{q/y}+\mu}{\omega_{q/y} -\omega_n} \right)$. This is the
term that
corresponds to the one which gives $f^{(1)}$
in (\ref{sum1}), see~\cite{lsscubic}. The new term comes from the simple pole at $p=-n$
as we have already pointed out. To  calculate this, one needs the
reflection identities a couple of times, and we find the contribution
\be 
\frac{A^{(1)}_{nq}e^{2\mu\tau_0}M(0^+)^2}{f^{(3)}_n n \omega_n
  \alpha_1 \alpha_2}.
\ee
Note that we choose a branch cut on $[ i\mu, -i\mu ]$, and the line integrals on either side of this cut
vanish since the integrand there is odd for $| { \rm Im}\, p|<\mu$. Finally,
the
asymptotics of the integrand are such that the integral on the large
circle tends to zero as the radius of the circle tends to infinity;
for this one needs the generalisation of Stirling's formula derived in~\cite{lsscubic}. 
Piecing all this together we deduce (\ref{lemma1}) for $r=1$. The $r=2$ version of the identity is easily
inferred from the $r=1$ by mapping $y \to 1-y$.

The second identity required is 
\begin{eqnarray}
\label{lemma2}
\sum_{q=1}^{\infty}\sum_{r=1}^3 \frac{\alpha_3}{\alpha_r}\,q \,
A^{(r)}_{mq}\, \frac{f^{(r)}_q}{\omega_{r,q} -\alpha_r \omega_n} =
\frac{(1+\mu\alpha k)}{\alpha_1 \alpha_2 \omega_m f^{(3)}_m}
\, \delta_{mn}.
\end{eqnarray}
This sum is very similar to another one evaluated in~\cite{lsscubic}, namely
(\ref{sum2}). The summand thus differs by a factor
of $(\omega_{r,q}+\mu\alpha_r)/(\omega_{r,q}-\alpha_r\omega_n)$, which
again does not change the asymptotics or the parity along the branch
cut of the corresponding integrand. Also note that the extra factor in
the numerator now ensures that the contribution from $q=0$ vanishes on
both sides of the branch cut. To prove (\ref{lemma2}) consider the
following contour integral
\begin{eqnarray}
\oint \frac{dq}{ 2\pi i} \frac{2}{\sin(\pi q)}\frac{(-1)^{m+1} \sqrt{m}
  \sin( m\pi y )}{q^2- y^2m^2} \frac{qM(0^+)}{\omega_{1,q}-y\omega_n}
  \frac{e^{\tau_0(\mu + \omega_{q/y})} \;
  \Gamma_{\mu(1-y)}(-q(1-y)/y)}{\omega_{q/y} \; \Gamma_{\mu}(-q/y)
  \Gamma_{\mu y}(q)}
\end{eqnarray}
where once again the contour is a large circle centred on the origin
which we will take to infinity. Note that now there is a branch cut on
$[ iy\mu, -iy\mu ]$. Now, the residues from the simple poles
at $q \in \mathbb{N}$ give us the $r=1$ term on the LHS of
(\ref{lemma2}). The simple poles at $q(1-y)/y \in \mathbb{N}$ give the
$r=2$ term, and the simple pole at $q=-my$ gives the $r=3$ term. The simple poles of $1/(\omega_{1,q}-y\omega_n$) at $q=ny$ and of
$1/(q^2-y^2m^2)$ at $q=my$, for $n \neq m$, get cancelled by the
factor $1/\Gamma_{\mu}(-q/y)$. For $n=m$ however, we have a simple
pole at $q=ny$. To evaluate its residue we need
\begin{eqnarray}
\Res_{q=ny} \left[
  \frac{1}{(\omega_{1,q}-y\omega_n)(q^2-y^2m^2)\Gamma_{\mu}(-q/y)}\right]=(-1)^{n+1}\frac{\omega_n}{2ny^2}\Gamma_{\mu}(n),
\end{eqnarray}
which after some manipulation, including the use of a reflection
identity, gives the following contribution to the above contour integral:
\be
-\frac{e^{2\tau_0\mu}M(0^+)^2}{\alpha_1\alpha_2\omega_n f^{(3)}_n}.
\ee
Observe that the line integrals along either side of the
branch cut for $|{ \rm Im} \, q |< y\mu$ vanish due to the integrand being
odd there. The circular integrals around the branch points also vanish as
$O(\epsilon^{1/2})$, see~\cite{lsscubic}, where $\epsilon$ is the radius
from the branch point. Finally, the asymptotics are such that the
integral on the circle at infinity vanish~\cite{lsscubic}.
This completes the proof of (\ref{lemma2}). 

Using (\ref{lemma1}) and (\ref{lemma2}), it is now a simple matter of some
algebra to check that the RHS of (\ref{check}) is equal to
$\delta_{mn}$. This completes the proof of existence of $\Gamma_+^{-1}$
and as we have already mentioned its uniqueness follows from the fact
that it must be both a left and right inverse as $\Gamma_+$ is a
symmetric matrix. 
\\

{ \it The author would like to thank Sakura Schafer-Nameki and Aninda Sinha
for reading through the manuscript and making helpful comments. The
author would also like to thank Malcolm Perry for encouragement to
publish the result. The author is supported by EPSRC.}

\section*{Appendix}
Here we summarise some useful definitions and identities.

\subsection*{Open-closed vertex}
Recall
the definitions of the two functions\footnote{Note that the definitions
appearing here are slightly different than in~\cite{lss}. They differ
in the denominators of the infinite product, and this only has the
effect of rescaling the Gamma functions by a $\mu$ dependent factor
(i.e. $\prod_{n=1}^{\infty} {\omega_{2n} \over 2n}$)
which cancels in the expressions for the Neumann matrices. This is to
simplify the reflection identities.} :
\begin{eqnarray}
\Gamma^{I}_{\mu} (z)
&=& e^{-\gamma \omega_{2z}/2}\left( {1\over z}\right) 
\prod_{n=1}^\infty \left({2n\over \omega_{2z}+\omega_{2n}} \,
e^{\omega_{2z}/ 2n}\right) \\
\Gamma^{II}_{\mu} (z)
&=&  e^{-\gamma (\omega_{2z-1}+1)/2} \left({2\over \omega_{2z-1}+ \omega_{1}}\right)
\prod_{n=1}^\infty \left({2n\over
\omega_{2z-1}+\omega_{2n+1}}\, e^{(\omega_{2z-1}+1)/ 2n}
\right) \,,
\end{eqnarray}
which satisfy the crucial reflection identities:
\begin{eqnarray}
\Gamma^I_{\mu}(z) \Gamma^I_{\mu}(-z) &=& -{\pi\over z \sin(\pi z)}  \\
\Gamma^{II}_{\mu}(1+z) \Gamma^{II}_{\mu}(-z)&=& -{\pi\over  \sin(\pi
z)}\,.
\end{eqnarray}
Note that both functions have simple poles for $z=-n$ where $n \in
\mathbb{N}$, and $\Gamma^{II}_{\mu}(z)$ also has a simple pole at
$z=0$. Also $\Gamma^I_{\mu}(z)$ has a branch cut on $[i\mu /2,
  -i\mu/2 ]$, whereas $\Gamma^{II}_{\mu}(z)$ has a branch cut on $[
  1/2+i\mu/2, 1/2-i\mu/2 ]$.

\subsection*{Cubic vertex}
Crucial quantities are the momenta of the three strings $\alpha_1$,
$\alpha_2$ and $\alpha_3$ which satisfy $\sum_{r=1}^{3} \alpha_r
=0$. We will always choose $\alpha_1 =y$, $\alpha_2=1-y$ and hence
$\alpha_3=-1$ as was done in~\cite{hssv, lsscubic}. Also $\tau_0=
\sum_{r=1}^{3} \alpha_r \log |\alpha_r|$ and $\alpha=\alpha_1\alpha_2\alpha_3$.
The matrices $A^{(r)}_{mn}$ and vector $B_m$ are given by:
\begin{eqnarray}
A^{(1)}_{mn} &=& { 2 \over \pi} (-1)^{m+n+1} \sqrt{mn} { \beta \sin(m \pi
\beta) \over n^2 - m^2 \beta^2} \cr 
A^{(2)}_{mn} &=& { 2 \over \pi} (-1)^{m+1} \sqrt{mn} { (\beta +1) \sin(m \pi
\beta) \over n^2 - m^2 (\beta+1)^2} \cr 
A^{(3)}_{mn} &=& \delta_{mn} \cr 
B_m &=& { 2 \over \pi}{ \alpha_3 \over \alpha_1 \alpha_2} (-1)^{m+1} {
\sin( m \pi \beta) \over m^{3/2}} \,,
\end{eqnarray}
where $\beta = \alpha_1/\alpha_3$. Another definition which we use is
\be
(U^{(r)})_{mn} = \delta_{mn} {(\omega_{r,m}-\alpha_r \mu)\over m}.
\ee
The functions $\Gamma_{\mu}^{(r)}(z)$ are defined as
\begin{eqnarray}
\Gamma^{(r)}_\mu (z) &=& e^{-\gamma \alpha_r \omega_z} \, {1\over \alpha_r z}\, \prod\limits_{n=1}^\infty
\left(
{n \over \omega_{r,n} + \alpha_r \omega_z} \;
e^{\alpha_r\omega_z\over n}\right) \\
&=& \Gamma^I_{2 \mu \alpha_r} (\alpha_r z)\,
\end{eqnarray}
and  $\Gamma_{\mu}(z) \equiv\Gamma^I_{2\mu}(z)$. They have simple
poles for $-z \in \mathbb{N}$ and branch cuts on $[i\mu, -i\mu]$. We have the
reflection identities
\be
\Gamma^{(r)}_{\mu}(z)  \Gamma^{(r)}_{\mu}(-z) = -{ \pi \over \alpha_r z \sin
(\pi \alpha_r z)}.
\ee
 The constant $M(0^+)$
is given by the function
\be
M(z)= {\Gamma_{\mu} (z) z \over \Gamma^{(1)}_{\mu} (z) \alpha_1 z \,
\Gamma_{\mu}^{(2)}(z) \alpha_2 z},
\ee
and $M(0^+)= \lim_{z \to 0^+} M(z)$.

\end{document}